\pdfoutput=1
\documentclass[10pt,conference]{IEEEtran} 

\IEEEoverridecommandlockouts
\usepackage{cite}
\usepackage{amsmath,amssymb,amsfonts}
\usepackage{algorithmic}
\usepackage{graphicx}
\usepackage{textcomp}
\usepackage{xcolor}
\usepackage{multirow}
\usepackage{booktabs}
\usepackage[most]{tcolorbox}
\usepackage{caption}
\usepackage{balance}
\graphicspath{ {./fig/}}
\usepackage[frozencache,cachedir=.]{minted}
\setminted[c]{
    fontsize=\footnotesize,
    xleftmargin=13pt,
    numbersep=5pt,
    linenos,
    texcomments,
    breaklines,
    frame=single,
    stripnl=false
}
\usepackage{hyperref}

\begin{document}
\title{Learning Program Semantics with Code Representations: An Empirical Study}

\author{
  \IEEEauthorblockN{%
    Jing Kai Siow\textsuperscript{1}\IEEEauthorrefmark{1}\,
    Shangqing Liu\textsuperscript{1}\IEEEauthorrefmark{1}\thanks{\IEEEauthorrefmark{1} Equal Contributions.}\,
    Xiaofei Xie\textsuperscript{2}\,
    Guozhu Meng\textsuperscript{3}\,
    Yang Liu\textsuperscript{4,1}\IEEEauthorrefmark{2}\thanks{\IEEEauthorrefmark{2} Corresponding author.}\,
  }
  \IEEEauthorblockA{%
    \textsuperscript{1}Nanyang Technological University, Singapore\\
    \textsuperscript{2}Singapore Management University, Singapore \\
    \textsuperscript{3}SKLOIS, Institute of Information Engineering, Chinese Academy of Sciences, China\\
    \textsuperscript{4}School of Information Science and Technology, Zhejiang Sci-Tech University, China \\
    \{jingkai001,shangqin001\}@e.ntu.edu.sg, xiaofei.xfxie@gmail.com, mengguozhu@iie.ac.cn, yangliu@ntu.edu.sg
  }
}

\maketitle
\begin{abstract}
Program semantics learning is the core and fundamental for various code intelligent tasks e.g., vulnerability detection, clone detection. A considerable amount of existing works propose diverse approaches to learn the program semantics for different tasks and these works have achieved state-of-the-art performance. However, currently, a comprehensive and systematic study on evaluating different program representation techniques across diverse tasks is still missed. 

From this starting point, in this paper, we conduct an empirical study to evaluate different program representation techniques. Specifically, we categorize current mainstream code representation techniques into four categories i.e., Feature-based, Sequence-based, Tree-based, and Graph-based program representation technique and evaluate its performance on three diverse and popular code intelligent tasks i.e., {Code Classification}, Vulnerability Detection, and Clone Detection on the public released benchmark. We further design three {research questions (RQs)} and conduct a comprehensive analysis to investigate the performance. By the extensive experimental results, we conclude that (1) The graph-based representation is superior to the other selected techniques across these tasks. (2) Compared with the node type information used in tree-based and graph-based representations, the node textual information is more critical to learning the program semantics. (3) Different tasks require the task-specific semantics to achieve their highest performance, however combining various program semantics from different dimensions such as control dependency, data dependency can still produce promising results. 
\end{abstract}

\section{Introduction}
\label{sec-intro}
With the booming development of open-source software, the amount of available code-related data has reached an unprecedented scale, which inspires researchers from both academia and the industry to explore employing data-driven approaches for diverse code-related problems such as type inference~\cite{type_inf, type_inf2, type_inf3}, clone detection~\cite{deckard, astnn, cc1}, source code summarization~\cite{liu2020retrieval, leclair2020improved, fernandes2018structured, ahmad2020transformer}, code search~\cite{gu2018deep, husain2019codesearchnet, shuai2020improving} and software vulnerability detection~\cite{devign, li2018vuldeepecker, russell2018automated}. Most of the existing works attempt to understand the behavior of the program i.e., understand the program semantics by the well-designed approaches for different tasks and have achieved promising results. Typically, we can broadly categorize these data-driven code-related works into four major categories: Feature-based, Sequence-based, Tree-based, and Graph-based representation to learn the program semantics for different tasks.

Feature-based approach requires the domain knowledge extracted from the program by the experts to represent the program semantics for different tasks. For instance, FLUCSS~\cite{fluccs} is designed for fault localization. It incorporated Spectrum Based Fault Localisation~\cite{survey} metric and source code related metrics to identify the fault in the software. It further performed learning-to-rank with Support Vector Machine~\cite{cortes1995support} and Genetic Programming~\cite{Vanneschi2012}. Their approach achieved a performance of 50\% Mean Average Precision (MAP) in ranking software defects. Bhel et al.~\cite{tfidf1} proposed mining security bugs from bug reports using TF-IDF~\cite{ref1} and Naive Bayes~\cite{nb} algorithm. They achieved high precision e.g., 92.56\% for detecting the bug report. 

Sequence-based representation treats code as a flat sequence of tokens~\cite{white_seq, machine_read} and converts them into numerical vectors with the distributed representation~\cite{word2vec} and further employs these vectors for diverse tasks. For example, VulDeepecker~\cite{vuldeepecker} predicted if a program is vulnerable by learning on code gadgets. Code gadgets are generated by slicing programs through the function calls. These code gadgets are then input into a bi-directional LSTM~\cite{bilstm} for learning whether the program is vulnerable. They achieved a score of 85.4\% in F1-score across multiple vulnerability types. Source code summarization is another popular application of the sequence-based approach. Hu et al.~\cite{ijcai2018-314} aimed to translate source code into their respective summary. They employed API sequences and code tokens as the input for an encoder-decoder model to generate the summary. They achieved 41.98 in BLEU-4~\cite{papineni-etal-2002-bleu} and 18.81 in METEOR~\cite{meteor}. 
\begin{figure*}
     \centering
     \includegraphics[width=1\textwidth]{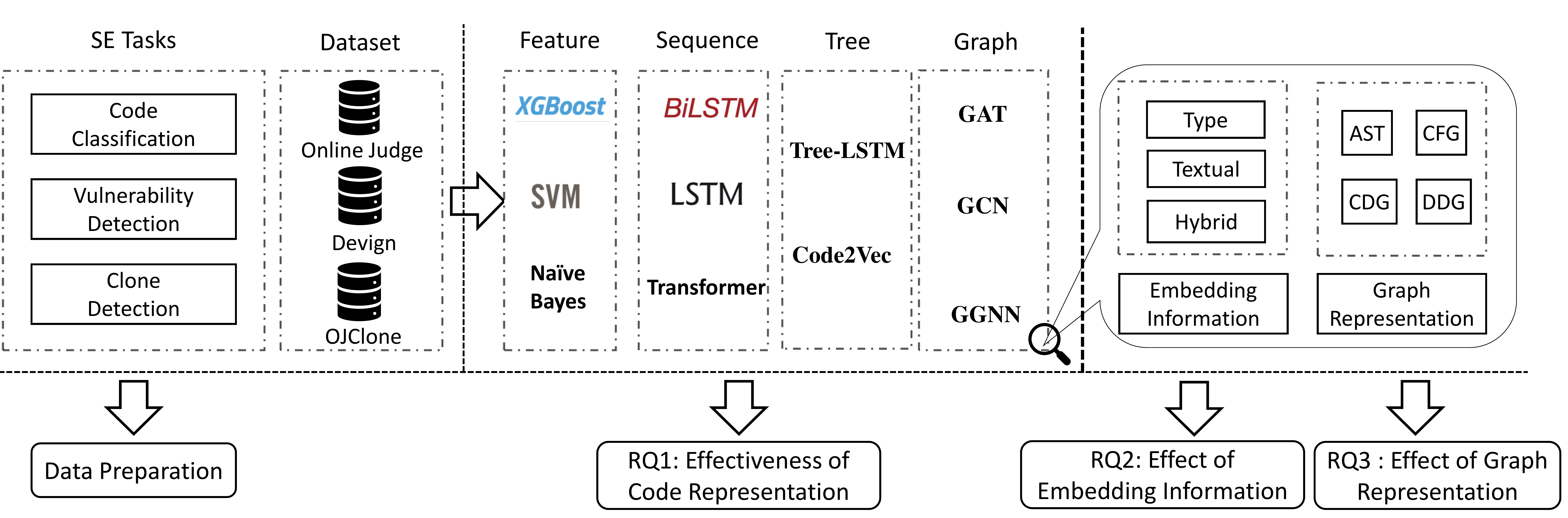}
     \caption{Overview of our empirical study.} 
     \label{fig:gnn_study}
\end{figure*}
Since programs are highly structured data, which can be converted into different structural representations such as the abstract syntax tree (AST). Many works attempt to explore this information hidden in the text for different tasks~\cite{code2seq, code2vec, bui2021treecaps, ijcai2017-0423}. For instance, code2vec~\cite{code2vec} predicted method name through the path contexts extracted from the AST of the program and achieved an F1-score of 58.4\%. CDLH~\cite{ijcai2017-0423} employed tree-based representation to detect code clones. It employed LSTM in learning an AST-based representation and hashed them to achieve a unique vector on each program. CDLH achieved promising results for Type-3 and Type-4 clone detection. 

The graph-based approaches attempt to embed more structural information such as data-flow, control-flow rather than the pure AST into a graph~\cite{learning_programs, gmn, liu2020retrieval, leclair2020improved} to represent the program semantics. For instance, Allamanis et al.~\cite{learning_programs} targeted at detecting variables that are incorrectly used in a project and predicting the correct variable. They enhanced AST with different data flow information by constructing diverse types of edges among the nodes, such as connecting variable nodes where the variable was last written to, to achieve a graph-based representation on the program. Their approach achieved a high accuracy of 85.5\% and 78.2\% in finding wrong variable names on the seen and unseen projects respectively. 

Despite different program representation techniques being widely utilized in learning program semantics for code-related tasks, currently, there is still a lack of a comprehensive study on evaluating and discussing the impact of different code representations across diverse tasks. Many challenges around code representation are still unsolved: (1) What type of the above code representation is optimal in the program scenario or in other words, whether there is an optimal code representation technique for different tasks? (2) The widely employed tree-based or graph-based approaches have shown superiority, however, these data usually contain complicated structures. For example, each node in AST usually contains node type and node textual information, whether they are both beneficial in learning program semantics? (3) The graph-based representation incorporates diverse program semantics such as data flow, control flow, data dependency, whether each component contributes equally to the final performance?
Based on the questions, we aim to answer the following research questions (RQs) as follows: 
\begin{itemize}
  \item \textbf{RQ1: Comparison of Feature-based, Sequence-based, Tree-based and Graph-based Representation.}
  \item \textbf{RQ2: Comparison of Node Embedding Information.}
  \item  \textbf{RQ3: Efficacy of Different Graph Representations.}
\end{itemize}
To answer these questions, we design our study on three popular and diverse code intelligent tasks: \textbf{Code Classification}, \textbf{Vulnerability Detection} and \textbf{Clone Detection} with typical approaches for these representations i.e., SVM, Naive Bayes, XGBoost for the feature-based representation; LSTM, BiLSTM, and Transformer for the sequence-based representation; Code2Vec, Tree-LSTM for the tree-based representation and Graph Convolution Neural Network (GCN), Graph Attention Network (GAT) and Gated Graph Neural Network (GGNN) for the graph-based representation. We further utilize the public released dataset for the evaluation and employ Joern~\cite{joern} for the unification on code property graph (CPG) construction for the graph-based representation. Specifically, CPG contains abstract syntax tree (AST), control flow graph (CFG), control dependency graph (CDG), and data dependency graph (DDG) to facilitate investigating each component. By the extensive experiments (around \textbf{1000} GPU hours), we conclude that: (1) Since graph-based representation incorporates diverse program semantics, it outperforms other compared representation techniques by a significant margin among the selected tasks. (2) Both node type and node textual information are beneficial in capturing program semantics, however, the textual information is more critical. (3) Different task relies on the task-specific semantics to achieve the best performance, however, generally, a composite graph representation with the comprehensive program semantics can still produce promising results. 

In summary, we make the following contributions:

\begin{itemize}
  \item To the best of our knowledge, this is the first large-scale empirical study that evaluates different code representation techniques on diverse popular evaluation tasks.
  
  \item We conduct comprehensive experiments and analysis to investigate the effect of each component in the tree and graph representation in capturing the program semantics on different tasks. 
    
  \item We provide extensive discussions from different aspects i.e., the learnt space by the different model, the semantic-preserving operation on the code snippet, and the statistical analysis of the program features on the predicted samples to illustrate the capacity and limitation of different representation techniques. We release our source code and data at \url{https://github.com/jingkai92/learning-program-representation} for the reproduction.
\end{itemize}

\begin{figure}
     \centering
     \includegraphics[width=0.5\textwidth]{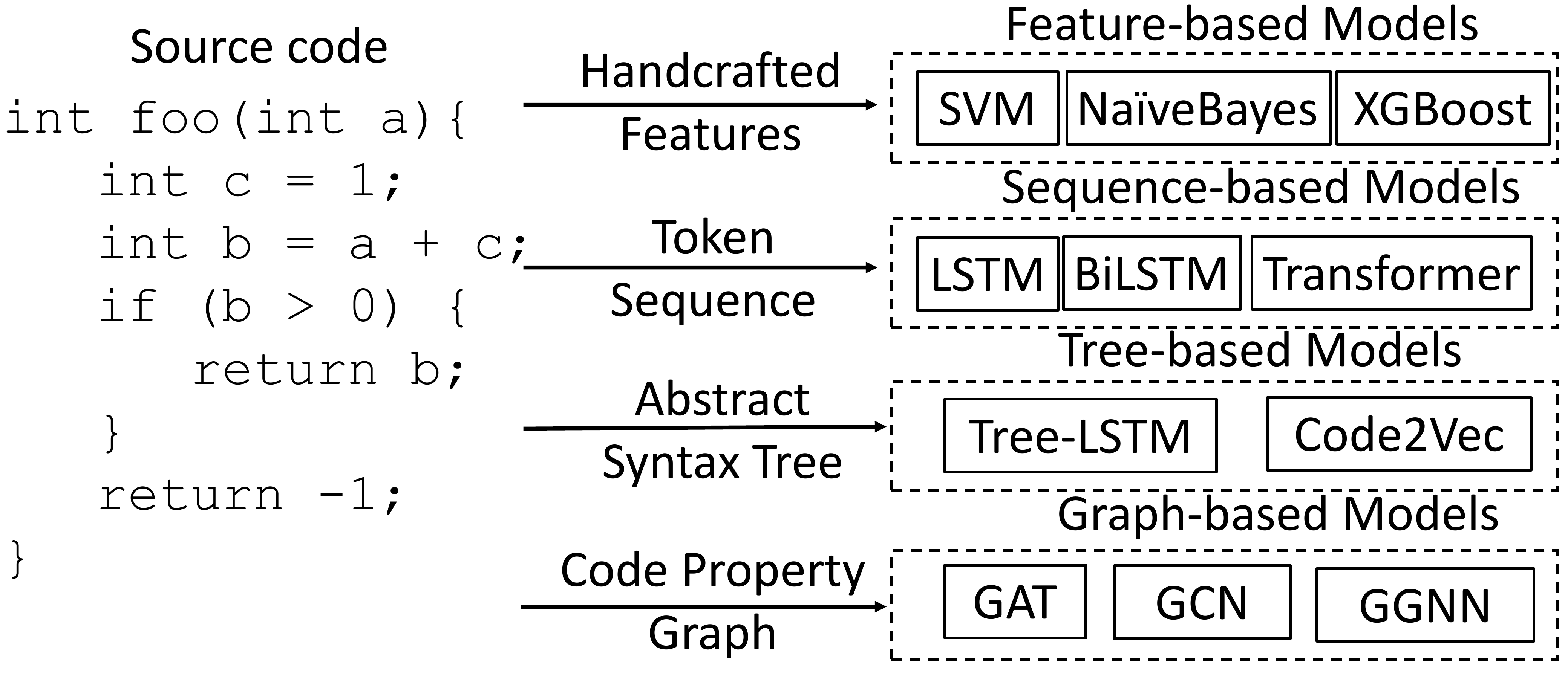}
     \caption{Techniques on Code Representation.} 
     \label{fig:code-reps}
\end{figure}

\section{Overview}
\label{sec-overview}
An overview of our study is shown in Fig.~\ref{fig:gnn_study}. In this study, we aim to answer the effectiveness of different code representations (Section ~\ref{sec:code-repr}) for different tasks (Section~\ref{sec:tasks}) on the public released benchmark. We categorize current state-of-the-art code representation techniques into feature-based, sequence-based, tree-based, and graph-based code representation and design RQ1-RQ3 (Section ~\ref{sec-emp}) for the investigation. 

\subsection{Evaluation Tasks}~\label{sec:tasks}
We select three diverse tasks i.e., code classification~\cite{astnn}, clone detection~\cite{astnn} and vulnerability detection~\cite{devign} for our study. We selected these tasks based on two reasons: (1) These tasks can effectively evaluate the semantic, lexical, syntactic information that is on the comparing evaluations. For instance, control dependency and data dependency are crucial in detecting vulnerabilities as the data flow of variables can indicate wrongful usage of variables~\cite{vuldeepecker}. Lexical information is important for clone detection as programs that uses the same tokens might be more likely to be cloned. A code representation that performs well in all three evaluating tasks infers that the representation can learn better semantic, lexical, and syntactic information. (2) The three evaluating tasks are popular in the software engineering domains. We collected the number of publications, in recent years in top leading conferences, e.g., ASE, ICSE, NeurIPS. We discovered that there are at least two, five, and four publications that are relevant to source code classification, vulnerability detection, and clone detection. This shows that our evaluating tasks are popular and important.
\begin{table}
\centering
\small
\caption{Statistics of Dataset.}
\label{tbl-dataset}
\begin{tabular}{c|ccc}
\toprule
\textbf{Dataset}                            & \multicolumn{1}{c}{} & \multicolumn{1}{c}{\textbf{\# of Samples}} & \multicolumn{1}{c}{\textbf{\# of Classes}} \\ \midrule
\multirow{3}{*}{Online Judge (OJ)} & Training             & 28622                          & \multirow{3}{*}{104}              \\
                         & Validation & 3581  &                    \\
                         & Testing    & 3628  &                    \\
                         \midrule
\multirow{3}{*}{Devign}  & Training   & 38526 & \multirow{3}{*}{2} \\
                         & Validation & 4815  &                    \\
                         & Testing    & 4817  &                    \\
                         \midrule
\multirow{3}{*}{OJClone} & Training   & 40000 & \multirow{3}{*}{2} \\
                         & Validation & 5000  &                    \\
                         & Testing    & 5000  &                   \\
                         \bottomrule
\end{tabular}
\end{table}

\noindent \textbf{Code Classification.} 
Code classification aims to classify the code fragments by their functionalities. which is vital for program understanding. Given a program, code classification aims to identify which category it belongs to from a category set. We employ OJ dataset~\cite{tbcnn}, a C Programming Language dataset, for code classification. The dataset contains 52,000 programs that are classified into 104 classes and each class performs a high-level operation, such as reversing an integer or finding minimum/maximum words from an array of words.

\noindent \textbf{Clone Detection.} 
Clone detection detects whether two code fragments achieve the same functionality with different implementations. We follow the same approach as Zhang et al.~\cite{astnn} and Wei et al.~\cite{ijcai2017-0423} to generate the code clone dataset based on OJ dataset~\cite{tbcnn}. Two programs in the same class can be considered as at least a functionality clone as they fulfill the same functionality~\cite{ijcai2017-0423}. Therefore, we gather the clones by pairing the programs that are in the same class and non-clones by pairing the programs that are in different classes. We randomly selected 25,000 programs and pair each of them with a clone and non-clones. We specifically ensure that a program that appears in the training set will not appear in the testing to ensure a fair evaluation. To differentiate from the dataset that we used in code classification, we refer to this dataset as OJClone.

\noindent \textbf{Vulnerability Detection.} 
The goal of vulnerability detection is to detect the vulnerable code fragments that may be attacked for cyber security. We utilize the completed Devign dataset~\cite{devign}, which contains 66,067 labeled functions that are collected from four open-sourced C programming language projects, FFmpeg, Wireshark, Linux, and Qemu for the evaluation. These functions are classified as either vulnerable or non-vulnerable. Zhou. et al. verify security patches manually and extract the functions from these verified patches~\cite{devign}, hence, their reliability is ensured. 

We remove the duplicate functions to ensure a fair comparison. Furthermore, we employ Joern~\cite{joern} to extract the tree and the graph from the programs. However, due to some compilation errors, finally, there is a set of 35,831 programs in our OJ dataset and 48,158 programs in the Devign dataset and the statistics of the datasets are shown in Table~\ref{tbl-dataset}. We split all our datasets into 80\% for the training set, 10\% for the validation set, and 10\% for the testing set.  

\begin{figure*}
     \centering
     \includegraphics[width=1\textwidth]{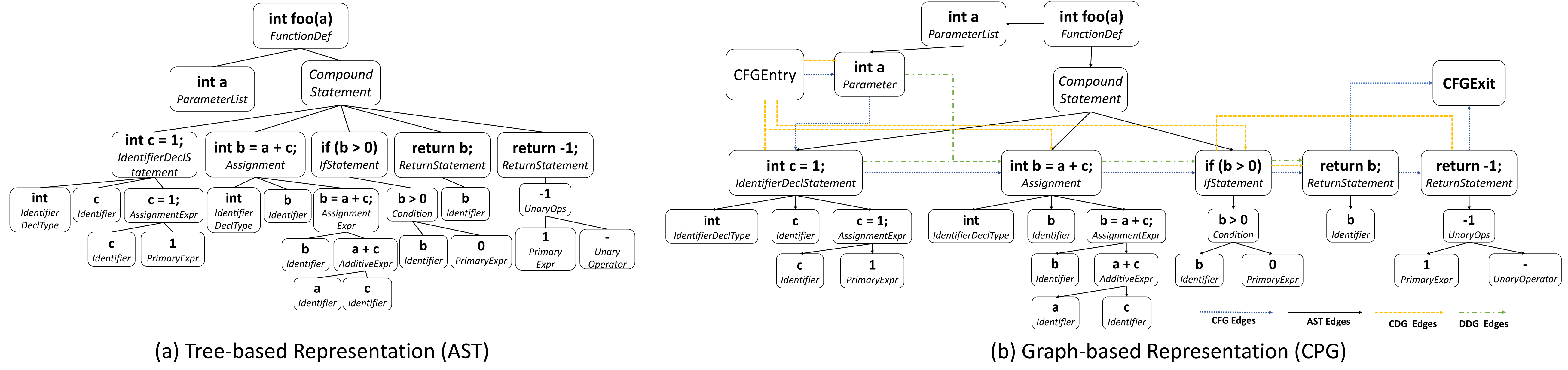}
     \caption{The Illustration of Tree and Graph Constructed by Joern where the original function is from Fig. \ref{fig:code-reps}.} 
     \label{fig:treegraph-reps}
\end{figure*}

\subsection{Code Representation}~\label{sec:code-repr}
As shown in Fig.~\ref{fig:code-reps}, to evaluate the efficacy of different code representation, we categorise the code representation into four major categories: Feature-based, Sequence-based, Tree-based and Graph-based Representation.

\noindent \textbf{Feature-based Representation.} In our study, we used Term Frequency-Inverse Document Frequency (TF-IDF) to vectorize the code snippet. TF-IDF computes a vector for each program based on the term frequency and inverse document frequency, which reduces the importance for stopwords and increases the weightage for more relevant words. To evaluate feature-based representation, we selected three machine learning algorithms: Support-Vector Machine (SVM)~\cite{cortes1995support}, Naive-Bayes~\cite{nb} and XGBoost~\cite{xgboost}. We use TF-IDF as our feature vector and input these vectors to the models. SVM performs the classification by finding a hyperplane in the dimensional space that can distinctly identify samples in different classes, while Naive-Bayes computes the probability of a sample within a class with an assumption of independent features. XGBoost is an implementation of a gradient boosting machine that is known for scalability and performance. These models are commonly used in many SE tasks as the comparison baselines~\cite{devign, astnn}. 

\noindent \textbf{Sequence-based Representation.} 
Since code snippets can be treated as a flat sequence of tokens, many works utilize this sequential information directly to learn program semantics. Following the previous works\cite{devign, allanamis_summarise,learning_programs}, we also employ token sequence directly in evaluating the sequence-based representation of the programs. We further employ sub-word splitting for efficient learning, where words are split by their camelcase and underscore. For instance, a function name, "get\_int", is split into {two subword tokens}, "get" and "int".  The result of this sub-word splitting is a sequence of subwords tokens. Their purpose is to reduce the vocabulary size of the datasets. For evaluating sequence representation, we opt to use Long-Short Term Memory (LSTM)~\cite{lstm} and Bi-directional LSTM (BiLSTM)~\cite{bilstm} network as they are suitable in training sequential data, i.e., sequence of tokens. For our experiment, we trained a LSTM and Bi-directional LSTM (BiLSTM) model to evaluate the performance on learning source code sequentially. We take the last hidden state of LSTM/BiLSTM as the learnt representation of the sequence representation. Furthermore, we added in transformer encoder~\cite{transformer} as part of our baseline to investigate the impact of the attention-based sequence approach. Transformer employs multi-head attention to learn a representation for a sequence and we utilize the contextual representation produced by the transformer encoder as the learnt program representation.    

\noindent \textbf{Tree-based Representation.} 
The program is a highly structured data compared {to} the sequential data, hence many code-related works attempt to extract the structure information e.g., abstract syntax tree (AST)~\cite{astnn, learning_programs, ijcai2017-0423, code2vec} behind the text to capture the semantics. {AST contains the syntactic information for a compiler to generate machine code, while removing unnecessary information, such as comments and whitespaces~\cite{dragon_book}.} An example of AST produced by Joern can be seen in Fig.~\ref{fig:treegraph-reps}(a). We selected two approaches for evaluating tree-based representations, Tree-LSTM~\cite{treelstm} and Code2Vec~\cite{code2vec}. Tree-LSTM employs LSTM in learning the network topology of the input tree structure, or in our case, the AST. It computes the hidden states based on its successors. As opposed to the single forget gate used in LSTM, it uses one forget gate for each child to focus on important information and outperforms several sequence-based approaches. Furthermore, several works~\cite{code2seq,shido2019} extend or employ Tree-LSTM as its baselines in software engineering. In our study, we employ the Child-Sum Tree-LSTM where the network learns the hidden states based on the summation of the children states. We employed max-pooling over all node representations to achieve the tree-based representation vector. Code2Vec~\cite{code2vec} is a state-of-the-art technique in code representation. It represents programs in a bag of path context, where path context represents a path between terminal nodes across the AST. Code representation is learnt through focusing on these path contexts. We extracted the path contexts from the tool ASTMiner~\cite{astminer} and adapted the source code given by Alon et al.~\cite{code2vec} for the evaluating tasks.  

\noindent \textbf{Graph-based Representation.} Many types of graphs are associated with programs, such as control flow graph (CFG), control dependency graph (CDG), data dependency graph (DDG).{Control flow graph depicts the execution flow of the statements in a program. Data dependency occurs when the value of a variable in a statement depends on the execution of the previous statement, whereas control dependency happens when the branching result of a predicate determines the execution of the immediate statements.} Existing works utilized these information for various tasks ~\cite{devign, liu2020retrieval, fernandes2018structured, leclair2020improved}.   
To investigate the impact of different types of graphs, we employ Code Property Graph (CPG) which is proposed by Yamaguchi et al.~\cite{cpg} by combining several graph representations, e.g., AST, CFG, CDG, and DDG, of source code into a graph structure. {We did not consider other graph structures such as Allamanis et al~\cite{learning_programs, fernandes2018structured} to represent program semantics as we follow the previous works~\cite{liu2020retrieval, devign}.} 
We selected three popular and widely used GNN variants i.e., Graph Convolutional Network (GCN), Graph Attention Network (GAT), and Gated Graph Neural Network (GGNN) to be our evaluating GNNs. These networks differ in their node message propagation, i.e., they aggregate and propagate information across the graph differently. GCN~\cite{kipf2017semi} uses first-order approximation of ChebNet~\cite{chebnet}. The neighboring information of a node is aggregated using convolutional operation and layers of networks can be stacked to enhance the learning of node features. Velickovic et al.~\cite{velickovic2018graph} propose GAT to use an attention mechanism in GNNs to attend over the neighborhood of a node to capture the local neighborhood information. GGNN~\cite{li2015gated} aggregates node information by Gated Recurrent Unit~\cite{gru} at every iteration to learn the node representations. For above GNN variants, to obtain the graph-level representation, which can be considered as the program representation for different tasks, we employ the max-pooling operation over the learnt node representations.

\noindent \textbf{Node Representation in Tree/Graph.} 
To obtain the tree structure or the graph structure for a program, we use Joern~\cite{joern}, a tool that is widely used academically~\cite{cpg} and commercially~\cite{joern}, to transform a function. A simple example of the source code in Fig.~\ref{fig:code-reps} is presented in Fig.~\ref{fig:treegraph-reps}, where Fig.~\ref{fig:treegraph-reps} (a) is the tree representation and Fig.~\ref{fig:treegraph-reps} (b) is the graph representation. We can find that each node has its node type, which distinguishes the node from the others (the second line in each node), and code textual information, which is a small fragment code snippet from the original program (the first line in each node). To get the initial node representation of the graph and tree, which is a vector uniquely to represent each node for the model learning, we employ three different embedding methods: {type embedding}, {textual embedding} and {hybrid embedding}. In {type} embedding, we embed each node solely by its node type, i.e., embed the node by the unique vectors that represent each type of node. Each node type is input into a linear layer to learn a unique representation. {Textual} embedding learns an intermediate representation of the node by inputting the textual information of the node into a BiLSTM. We use the last hidden state of the BiLSTM as our initial node representation for the embedding. For the node without textual information such as the compound statement node, we use the empty string for the replacement and feed the empty string into BiLSTM to get the representation. {Hybrid} embedding employs a linear layer to learn the concatenated representation of textual embedding and type embedding. Then these embedded node representations are utilized with Tree-LSTM or GNNs respectively to learn the representations.

\section{Empirical Study}
\label{sec-emp}
In this section, we detail our experimental settings to ensure transparency in our experiments. We then answer the proposed RQ1-RQ3 with a comprehensive analysis. 

\subsection{Experiment Settings} \label{evaluation_metrics}

\noindent \textbf{Experimental Setup.} We embed each function into TF-IDF feature vector and input these vectors into the SVM, Naive Bayes, and XGBoost, resulting in a list of class probabilities. The class with the highest probability will be our prediction for the model. We employ Multinomial Naive Bayes~\cite{nb} and Radial Basis Function Kernel for SVM~
\cite{cortes1995support}. For XGBoost~\cite{xgboost}, we trained our model with 40 rounds with a tree-depth of 32. We used the following hyper-parameters for LSTM, BiLSTM, and Transformer network: 128 for word embedding dimension and LSTM hidden size, 4 layers of LSTM/BiLSTM and transformer encoder, 4 heads for transformer attention layer. We used DGL~\cite{dgl} as our implementation for Tree-based and Graph-based approaches. Similarly, we used a hidden dimension of 128. For GCN, GAT, and GGNN, the following hyperparameters are used: 128 for word embedding and initial node representation, 4 layers for GAT/GCN/GGNN, and 4 attention heads for GAT. We train all our baselines with training data and tune the models based on the validation data. We then report the performance of the models using the testing data. To ensure fairness in the evaluation, we fixed all the dimensions to be 128 and uses a learning rate between 0.01 to 0.0001 for different tasks, batch size of 128, and a dropout rate of 0.2. All models are trained until 50 epoch and early stopping of 10 epoch. We limit the number of tokens in the function to be 150 and restrict the size of the graph to be within 250 nodes. For all deep learning experiments, we employ an additional linear classifier to learn on classifying the learnt representation into their respective classes. All experiments are conducted on an Intel(R) Xeon(R) CPU E5-2698 v4 @ 2.20GHz Linux 16.04 server with 128GB RAM and equipped with three Tesla V100-SXM2-32GB graphic cards.

\begin{table*}
\centering
\caption{Results of Code Classification, Vulnerability Detection and Clone Detection.}
\label{tbl-result-code-repr}
\begin{tabular}{l|c|cc|cc}
\toprule
\multicolumn{1}{c|}{\multirow{2}{*}{\textbf{Models}}} &
  \textbf{\begin{tabular}[c]{@{}c@{}}Code \\ Classification\end{tabular}} &
  \multicolumn{2}{c|}{\textbf{\begin{tabular}[c]{@{}c@{}}Vulnerability \\ Detection\end{tabular}}} &
  \multicolumn{2}{c}{\textbf{\begin{tabular}[c]{@{}c@{}}Clone \\ Detection\end{tabular}}} \\
\multicolumn{1}{c|}{} & Accuracy & Accuracy & F1     & Accuracy & F1    \\ \midrule
SVM                           & 0.5413 & 0.5223 & 0.5144 & 0.6631 & 0.6780 \\
Naive Bayes                 & 0.2762 & 0.6934 & 0.6762 & 0.5493 & 0.6330 \\
XGBoost                       & 0.5929 & 0.7056 & 0.6951 & 0.7773 & 0.7979 \\ \midrule
LSTM                          & 0.8094 & 0.7098 & 0.7135 & 0.8298 & 0.8414 \\
BiLSTM                        & 0.8382 & 0.7131 & 0.7162 & 0.8502 & 0.8544 \\
Transformer Encoder           & 0.8193 & 0.4796 & 0.6482 & 0.5000 & 0.6660 \\ \midrule
Code2Vec                      & 0.8973 & 0.7180 & 0.7192 & 0.6180 & 0.6719 \\
Tree-LSTM           & 0.8600 & 0.7100 & 0.7209 & 0.9024 & 0.9055 \\ 
\midrule
GCN                 & 0.8936 & 0.7015 & 0.7289 & 0.9166 & 0.9188 \\
GAT               & 0.9042 & \textbf{0.7278} & 0.7306 & 0.8982 & 0.8997 \\
GGNN               & \textbf{0.9204} & 0.7158 & \textbf{0.7344} & \textbf{0.9350} & \textbf{0.9367}\\ 
\bottomrule

\end{tabular}
\end{table*}

\noindent \textbf{Evaluation Metrics.} We adopt accuracy and F1 as the evaluating metrics. Accuracy computes the correct prediction of each class and averages it with the total number of samples. F1 is commonly used in binary classification~\cite{astnn, devign, ijcai2017-0423}. It is computed using a weighted combination of precision and recall. A high F1 implies the model has a low number of false positives and false negatives. The higher values of these metrics, the better performance the approach achieves. 

\subsection{RQ1: Comparison of Different Code Representations}

Table~\ref{tbl-result-code-repr} shows the comparison results of our experiments. For this RQ, we employed hybrid embedding for tree representation and graph representation to learn the programs. We observed that among all the feature-based models, XGBoost performs the best across all three tasks, in contrast, Naive Bayes has the worst performance since the dependencies among the tokens in the program cannot be captured. 

From the third row of Table~\ref{tbl-result-code-repr}, we can find that the sequence-based models have a better performance as compared to the feature-based approaches and BiLSTM performs the best across all three tasks, having accuracy in 83.82\% in Code Classification, and F1-Score of 71.62\% in Vulnerability Detection and accuracy of 85.44\% in Clone Detection. This indicates that learning dependency among the tokens in the program is important for learning program semantics and these tokens have the rich semantic information to represent the programs, which yields better performance compared with the feature-based approaches. 

The tree-based approaches such as Tree-LSTM has the better performance over these tasks compared with the sequence-based approaches. Specifically, Tree-LSTM has an improvement of 2.18\%, 0.47\%, and 5.11\% in accuracy and F1-Score over BiLSTM, which shows that the semantic and syntactic information on AST can be useful in learning the representation for the program, An interesting finding is that Code2Vec performs better than BiLSTM in both code classification and vulnerability detection, with an increase of 5.91\% in code classification accuracy and 0.3\% in F1-Score of vulnerability detection. However, the performance in clone detection is lacking. We attribute the low performance to the different purposes of the code representation. The original purpose of code2vec is to predict method names in Java programs~\cite{code2vec}. Hence, path contexts might be better for predicting function names but lacks in detecting clones.   

Finally, we can observe that the graph-based approaches perform the best on the evaluating tasks compared with other representations. 
GGNN outperforms all other non-graph representations by 2.31-64.42\% in code classification accuracy, {1.35\%-22.0\% in F1-Score of vulnerability detection, and 3.12-30.37\% in the accuracy of clone detection.} Furthermore, the difference between the GNN variants such as GCN, GAT, and GGNN is not very obvious for example, GGNN and GAT have only a difference of 1.62\% in code classification accuracy, 0.38\% in vulnerability detection F1-Score, and 3.7\% in clone detection F1-Score. We infer that since graph representation embeds more semantics of the programs compared with other baselines, hence it outperforms other baselines by a significant margin. However, the impact between different variants of GNNs is minor. 

\begin{tcolorbox}[breakable,width=\linewidth,boxrule=0pt,top=1pt, bottom=1pt, left=1pt,right=1pt, colback=gray!20,colframe=gray!20]
\textbf{Answer to RQ1:} Graph-based representations are best in representing program semantic among all our comparing representations. We achieve improvements up to 64.42\% accuracy in code classification, 8.62\% F1-Score in clone detection, and 30.37\% F1-Score in vulnerability detection when graph-based representations are used.

\textbf{Insights:} Graph-based representation is superior to the sequence-based or tree-based representation for many tasks. However, the construction of the graph for the program is non-trivial which requires extra efforts and this limits the usage of graph-based representation. It is crucial to have better tools to facilitate the code property graph construction for other programming languages such as Java and Python. 
\end{tcolorbox}

\begin{table*}
\centering
\caption{Results of Embedding Information.}
\label{tbl-result-node-repr}
\begin{tabular}{l|c|cc|cc}
\toprule
\multicolumn{1}{c|}{\multirow{2}{*}{\textbf{Embedding}}} &
  \textbf{\begin{tabular}[c|]{@{}c@{}}Code \\ Classification\end{tabular}} &
  \multicolumn{2}{c|}{\textbf{\begin{tabular}[c]{@{}c@{}}Vulnerability \\ Detection\end{tabular}}} &
  \multicolumn{2}{c}{\textbf{\begin{tabular}[c]{@{}c@{}}Clone \\ Detection\end{tabular}}} \\ 
\multicolumn{1}{c|}{}  & Accuracy & Accuracy & F1     & Accuracy &  F1 \\ \midrule
Tree-LSTM (Type) & 0.7657   & 0.6207   & 0.6507 & 0.7862   & 0.8085     \\
GCN (Type)       & 0.8198   & 0.6101   & 0.6300 & 0.8410   & 0.8470     \\
GAT (Type)       & 0.8860   & 0.5277   & 0.6567 & 0.7972   & 0.8067     \\
GGNN (Type)      & 0.8787   & 0.5128   & 0.6602 & 0.8508   & 0.8558     \\ \midrule
Tree-LSTM (Textual)   & 0.8380   & 0.7162 &	0.7092 & 0.8606   & 0.8726     \\
GCN (Textual)         & 0.8503   & 0.7183   & 0.7184 & 0.9036   & 0.9055     \\
GAT (Textual)         & 0.8930   & \textbf{0.7289} & 0.7153 & 0.8906 & 0.8945     \\
GGNN (Textual)        & 0.8839   & 0.7233 & \textbf{0.7362} & 0.9094 & 0.9116    \\ \midrule
Tree-LSTM  (Hybrid)          & 0.8600 & 0.7100 & 0.7209 & 0.9024 & 0.9055 \\
GCN (Hybrid)                 & 0.8936 & 0.7015 & 0.7289 & 0.9166 & 0.9188 \\
GAT (Hybrid)                 & 0.9042 & 0.7278 & 0.7306 & 0.8982 & 0.8997 \\
GGNN (Hybrid)               & \textbf{0.9204} & 0.7158 & 0.7344 & \textbf{0.9350} & \textbf{0.9367}\\ 
\bottomrule
\end{tabular}
\end{table*}

\subsection{RQ2: Comparison of Different Node Embedding Information}
In this RQ, we want to investigate the impact of different information that is embedded into the node representation for tree-based or graph-based representation. Specifically, we ablate the performance that is embedded with type, textual, and hybrid embedding. Table~\ref{tbl-result-node-repr} shows the results. 

We observe that compared with the type embedding, textual embedding has a significant improvement. Specifically, the improvement in code classification accuracy ranges from 0.07\% (GAT) to 7.23\% (Tree-LSTM), in vulnerability detection F1-Score in a range of 5.85\% (Tree-LSTM) to 8.84\% (GCN), and in clone detection F1-Score in a range of 5.58\% (GGNN) to 8.78\% (GAT). We claim it is reasonable since the tokens tend to carry more semantic information of the program, {e.g.,} a program of the function name \textit{reverse\_array}, which we can infer that it is to finish a reversal operation on the input array, and ignore this textual information will increase the difficulty of the model to capture the functionality. Furthermore, we also perform a simple statistic analysis on the OJ dataset for the code classification, we find that the average Jaccard Index (a metric to measure the text-similarity) for programs is 0.51, which indicates there are many overlap tokens between the programs. Hence, ignoring the token information in the node will harm the performance significantly. 

Furthermore, we can see that combining textual and type embedding i.e., hybrid embedding, will further improve the performance over code classification and clone detection tasks. For instance, the performance of GGNN and Tree-LSTM increases by 3.65\%, 2.20\% when hybrid embedding is used on the code classification. We infer that incorporating both node type and node textual information can improve the model capacity and brings better performance. However, on the vulnerability detection, using type embedding has a negative impact. We conjecture that it is due to the way of combining embeddings. In the hybrid embedding, we just employ a linear layer to concatenate the representation of the textual embedding and type embedding, which is simple and straightforward. Vulnerability detection is a much-complicated task, especially for the real vulnerabilities. Although hybrid embedding obtains sub-optimal performance on the vulnerability detection, however, on the other tasks, it still gets the best performance. We will explore a more effective combining way and leave it as our future work.

\begin{tcolorbox}[breakable,width=\linewidth,boxrule=0pt,top=1pt, bottom=1pt, left=1pt,right=1pt, colback=gray!20,colframe=gray!20]
\textbf{Answer to RQ2:} 
Textual information is more critical to learning the program semantics for the tasks as compared to the node type for the tree-based and graph-based representation. Furthermore, combining both with a simple linear layer, we can obtain the optimal performance on code classification and clone detection and sub-optimal performance on vulnerability detection.

\textbf{Insights:} 
The combination way i.e., a single linear layer is simple and straightforward. Although it generates a promising performance on code classification and clone detection, more tasks need to evaluate its generalization. Furthermore, it is also valuable to explore some other combination ways.
\end{tcolorbox}

\begin{table}
\caption{Results of Graph Representation Analysis with GGNN.}
\label{tbl-result-edge-analysis}
\begin{tabular}{l|c|cc|cc}
\toprule
\multicolumn{1}{c|}{\multirow{2}{*}{\textbf{Models}}} &
  \textbf{\begin{tabular}[c]{@{}c@{}}Code \\ Classification\end{tabular}} &
  \multicolumn{2}{c|}{\textbf{\begin{tabular}[c]{@{}c@{}}Vulnerability \\ Detection\end{tabular}}} &
  \multicolumn{2}{c}{\textbf{\begin{tabular}[c]{@{}c@{}}Clone \\ Detection\end{tabular}}} \\
\multicolumn{1}{c|}{} & Accuracy & Accuracy & F1 & Accuracy & F1 \\ \midrule
AST & 0.8734 & 0.7033 & 0.7125 & 0.9172 & 0.9204 \\
CFG & 0.8890 & 0.7042 & 0.7085 & 0.9276 & 0.9300 \\
CDG & 0.8856 & 0.7160 & 0.7120 & 0.9144 & 0.9176 \\
DDG & 0.8339 & \textbf{0.7235} & 0.7133 & 0.9222 & 0.9251 \\ 
\midrule
CPG & \textbf{0.9204} & 0.7158 & \textbf{0.7344} & \textbf{0.9350} & \textbf{0.9367} \\ 
\bottomrule
\end{tabular}
\end{table}

\subsection{RQ3: Efficacy of Different Graph Representations}
In this RQ, we study the impact of different graph representations on the performance of the evaluating tasks. Specifically, we evaluate the performance of AST, CFG, CDG, and DDG across the three evaluating tasks. Among all comparing graph neural networks, GGNN has the highest performance as shown in RQ1 and RQ2. Hence, we employed GGNN for this evaluation. 

The results are shown in Table~\ref{tbl-result-edge-analysis}. We observe that CFG performs better than other graph representations in code classification and clone detection, however on the vulnerability detection, DDG can achieve higher performance. We believe it is reasonable, as data dependency is easy to trigger the vulnerabilities since vulnerable functions tend to have complex dependencies among the statements, e.g., memory de-referencing, and buffer overflows. Furthermore, according to Fabian et al. ~\cite{yamaguchi2015pattern}, it is inherent that DDG has a bigger influence on finding vulnerable functions. We show an example for the illustration. 
Fig~\ref{fig:vuln-rq3} shows an example of a function~\footnote{\href{https://github.com/FFmpeg/FFmpeg/commit/f42b3195d3f2692a4dfc0a8668bb4ac35301f2ed}{A commit with its commit id (f42b31).}} from FFmpeg, which corresponds to a out-of-bound (OOB) bug. GGNN predicted the correct label for this function with the DDG representation, which is shown in Fig~\ref{fig:ddg}. Specifically, we can see that the ``buffer'' array keeps increasing by adding the variable ``i'' without OOB guard. This results in the indexing of the array possibly growing out of the array size. In DDG, the relationship between the increment of ``i'' and the ``buffer'' indexing can be directly observed and there is a dependency between the nodes with the types ``ArrayIndexing'' and ``IncDecOp''. However, this relationship is missed in other representations such as AST and CFG. 

\begin{figure}[t]
    \begin{minipage}[t]{0.45\textwidth}
        \centering
    \caption{Vulnerable function that required data dependency for the detection.}
    \label{fig:vuln-rq3}
    \begin{minted}{c}
    static void fix_bitshift(ShortenContext *s, int32_t *buffer)
    {
        int i;
        if (s->bitshift != 0)
            for (i = 0; i < s->blocksize; i++)
                buffer[s->nwrap + i] <<= s->bitshift;
    }
    \end{minted}
    \end{minipage}
\end{figure}

\begin{figure}
	\centering
	\includegraphics[width=60mm,scale=0.3]{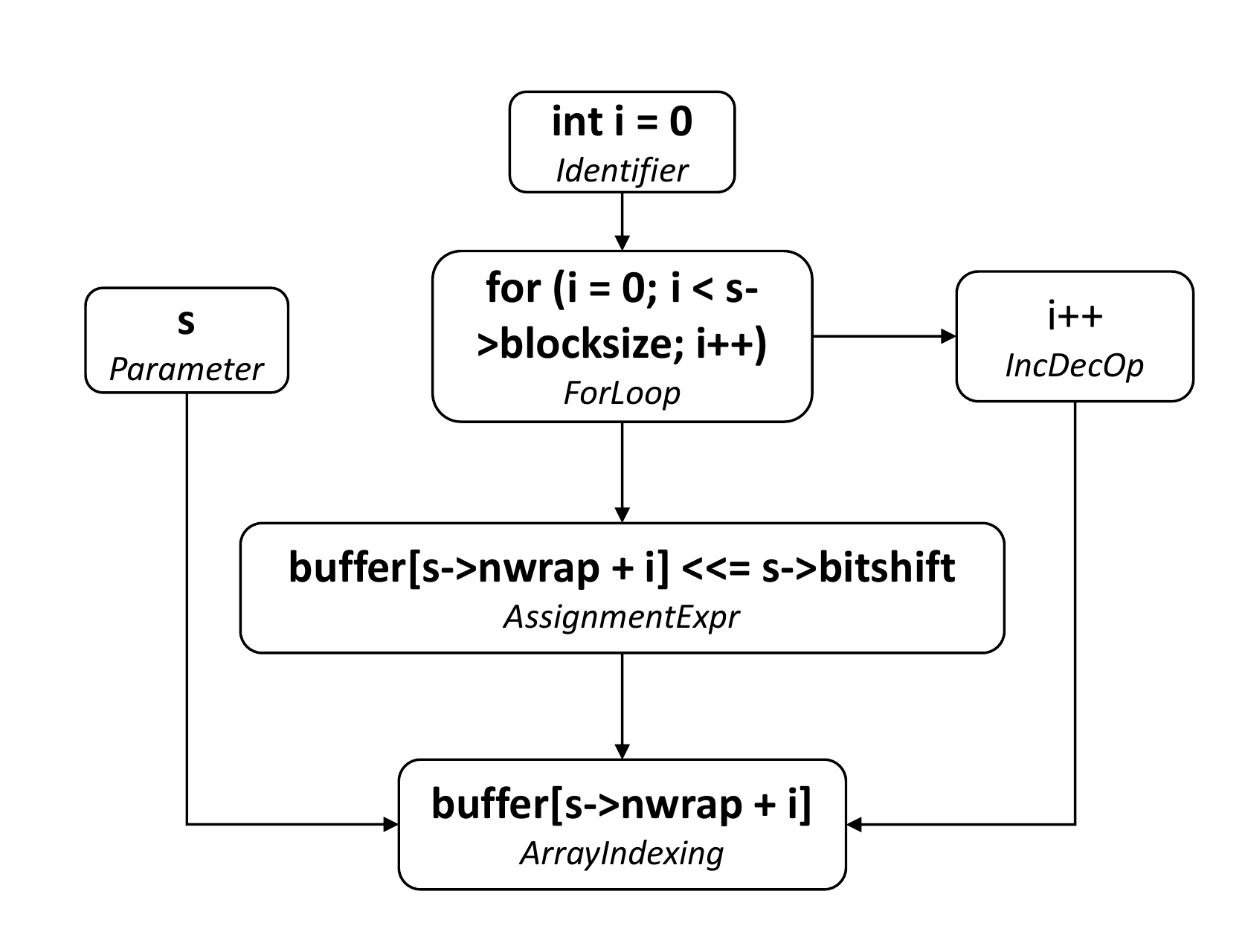}
	\caption{Data Dependency Graph of the function in Fig~\ref{fig:vuln-rq3}.} 
	\label{fig:ddg}
\end{figure}

\begin{figure*}
	\centering
	\includegraphics[width=150mm,scale=0.3]{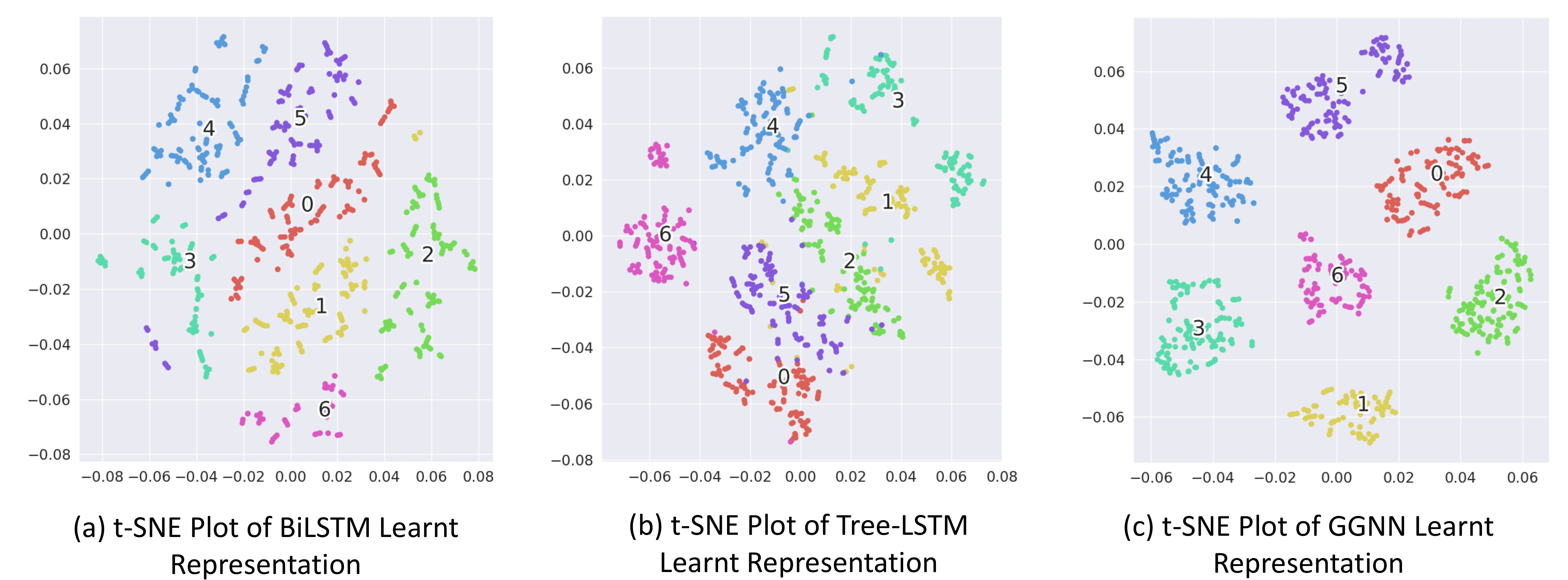}
	\caption{t-SNE plot of Learnt Representation Space.} 
	\label{fig:tsne}
\end{figure*}

Furthermore, when all graph representations are combined i.e., CPG, it achieves better performance on code classification and clone detection and yields the sub-optimal results on vulnerability detection i.e., the accuracy of CPG (0.7158) is lower than DDG (0.7235), but F1-Score is still the highest (0.7344). It indicates that due to the different characteristics of the task, there might be some specific semantics that are particularly suitable for this task e.g., the data dependency for vulnerability detection and produce the best performance. However, a composite graph with immense semantic and syntactic information can still achieve promising results.

\begin{tcolorbox}[breakable,width=\linewidth,boxrule=0pt,top=1pt, bottom=1pt, left=1pt,right=1pt, colback=gray!20,colframe=gray!20]
\textbf{Answer to RQ3:}  Different task relies on the task-specific semantics to achieve the best performance, however, generally speaking, a composite graph with the comprehensive program semantics can yield the promising results.

\textbf{Insights:}
Combining diverse dimensional code semantics, e.g., AST, CFG, CDG, DDG are beneficial for the neural networks to capture the program semantics, However,
from one perspective, how to capture the task-specific semantics for a task to achieve the best results is still an open-question, from another perspective, CPG only contains syntactic information (AST), data flow information (CDG, DDG), control flow information (CFG), hence, is it sufficient to represent program semantics? 
\end{tcolorbox}

\section{Discussion}
\label{sec-disc}
In this section, we first investigate the learnt space by the selected models, then perform an experiment on the semantic-preserving transformation to explore the capacity of different models and conduct a statistic analysis on the predicted results. Finally, we present the threats to the validity of this work.

\begin{figure*}[!htb]
    \begin{minipage}[t]{0.45\textwidth}
        \centering
    \caption{Example 1 of Vulnerable Function with its transformed version.}
    \label{fig:masked-vuln-1}
    \begin{minted}{c}
    /* Original Function */
    static inline void put_codeword(PutBitContext *pb, vorbis_enc_codebook *cb, int entry)
    {
        assert(entry >= 0);
        assert(entry < cb->nentries);
        assert(cb->lens[entry]);
        put_bits(pb, cb->lens[entry], cb->codewords[entry]);
    }
    
    /* Transformed Function */
    static inline void f1(v2 *pb, v2 *v1, int entry)
    {
        assert(v1->lens[entry]);
        assert(entry < v1->nentries);
        assert(entry >= 0);
        put_bits(pb, v1->lens[entry], v1->codewords[entry]);
    }
    
    \end{minted}
    \end{minipage} \hspace{0.45 in}
    \begin{minipage}[t]{0.46\textwidth}
        \centering
    \caption{Example 2 of Vulnerable Function with its transformed version.}
    \label{fig:masked-vuln-2}
    \begin{minted}{c}
    /* Original Function */
    static void pic_common_class_init(ObjectClass *klass, void *data)
    {
        DeviceClass *dc = DEVICE_CLASS(klass);
        dc->vmsd = &vmstate_pic_common;
        dc->no_user = 1;
        dc->props = pic_properties_common;
        dc->realize = pic_common_realize;
    }
    
    
    /* Transformed Function */
    static void p1(ObjectClass *k1, void *d1)
    {
        DeviceClass *d1 = DEVICE_CLASS(k1);
        d1->realize = pic_common_realize;
        d1->no_user = 1;
        d1->vmsd = &vmstate_pic_common;
        d1->props = pic_properties_common;
    }
    \end{minted}
    \end{minipage} \\
\end{figure*}

\subsection{Learnt Representation Space by Neural Network}

To demonstrate the learning capability of GGNN over BiLSTM and Tree-LSTM, we employ t-SNE~\cite{maaten2008visualizing} to visualize the learnt representation space of code classification. Specifically, we randomly picked 7 classes in the testing dataset. The learnt space is shown in Figure~\ref{fig:tsne}, where the class is labeled along with the color in the plot. We can observe that the graph representation performs the best. As shown in Fig.~\ref{fig:tsne}(a) and Figure~\ref{fig:tsne}(b), the representations learnt by BiLSTM are more scattered across the plot than the learnt space by Tree-LSTM, which means that the distances of any samples from the same class are greater in BiLSTM. Hence, compared with BiLST, Tree-LSTM produces a better learning space. The learnt space of GGNN is shown in Figure~\ref{fig:tsne}(c), we can easily find that the boundary for each class is more clear and the aggregated cluster is more condensed compared with Tree-LSTM. This indicates that GGNN has a more powerful learning capacity compared with BiLSTM and Tree-LSTM. 

\subsection{Semantic-Preserving Transformation on Code Representation}~\label{sec-quali} 
We further explore the capacity of these models on semantic-preserving operations. Here the semantic-preserving operation is defined to transform the code snippet with simple operations such as renaming the identifiers or swapping two independent statements while keeping the original program semantics. Specifically, we randomly select two vulnerable functions~\footnote{{Both functions are from FFmpeg(\href{https://github.com/FFmpeg/FFmpeg/commit/1ba08c94f5bb4d1c3c2d3651b5e01edb4ce172e2}{1ba08c}) and Qemu(\href{https://github.com/qemu/qemu/commit/efec3dd631d94160288392721a5f9c39e50fb2bc}{efec3d}) respectively.}} that BiLSTM, Tree-LSTM and GGNN predict correctly from the {Devign} test set and both examples are shown in Fig.~\ref{fig:masked-vuln-1} and Fig.~\ref{fig:masked-vuln-2} respectively where the top section shows the original function, while the bottom section are the transformed version. {The original function in Fig.~\ref{fig:masked-vuln-1} lacks a buffer overwrite protection, hence is exposed to buffer overflow vulnerability. For the function in Fig.~\ref{fig:masked-vuln-2}, a deprecated variable, ``dc-\textgreater no\_user'' is used. This might cause a regression bug where a previously working version stops working.} We conduct two kinds of simple transformation operations: (1) We randomly swap the location of two statements, that are independent with others. For instance, in Fig.~\ref{fig:masked-vuln-1}, we swap Line 4 and Line 6 in the original function. This does not affect any dependency as they are independent assertion statements. (2) We randomly select some variables and replace all occurrences of the variable name into the meaningless placeholder. We select both operations since they are simple and easy for implementation.

We input the transformed functions with the trained BiLSTM, Tree-LSTM, and GGNN respectively to investigate the performance. Surprisingly, we find that TreeLSTM and GGNN can produce the correct prediction, while BiLSTM fails. We infer that since the sequence-based representations model the permutation and sequential information of the statements to capture the semantics, hence they are more susceptible to the ``swap'' and ``renaming'' operations while these operators do not destruct the original semantics. In contrast, due to Tree-LSTM and GGNN both employ the structure information to capture the semantics, they are more robust to these simple transformations and hence produce the correct predictions.  

\subsection{Analysis of Prediction Results} \label{sec-incorrect}
By Table~\ref{tbl-result-code-repr}, we have proved that the graph-based representation has the best performance across three tasks, to further analyze its capacity, we conduct the statistical analysis on the specific complex features in the program that GGNN cannot learn well for the code classification. Specifically, we manually examine the correct and incorrect predicted programs on the test set in code classification. We utilize the correct predicted programs (total 149) from the top-three best performing classes (Class 29, Class 66, Class 83) and incorrectly predicted programs (total 63) from the top-three worst-performing classes (Class 25, Class 52, Class 61) for investigation. We further summarize 5 complex features that may be existed in the program: Nested Loops, Multiple Loops, Pointer Operation, Do-While, and Others/Basic. Multiple loops refer to the loops that are initialized in different scopes and nested loops refer to the loops that are initialized within another loop in a program. We group programs under Pointer Operation whenever de-referencing of address occurs in the program. Do-while is an alternative type of looping mechanism. Lastly, if a program has no above-defined complex characteristics e.g., a program that only has single loops and assignment statements, we group it into the Others/Basic category.

\begin{table}[t]
\small
  \centering
  \addtolength{\tabcolsep}{2pt}
	\caption{Complex Features in Classified Programs.}
	\label{tbl-complex-features}
	\scalebox{0.9}{
	\begin{tabular}{c | c c | c c }
	
	    \toprule 
        \multirow{2}{*}{\textbf{Complex Features}} & \multicolumn{2}{c|}{Incorrect Predicted} &    \multicolumn{2}{c}{Correct Predicted}  \\ 
	    
	    & Num & Ratio  & Num & Ratio  \\ 
	    
	    \midrule
	    Do-While           & 0   & 0\%      & 3   &  1.84\% \\
	    Multiple Loops     & 29  & 34.11\%  & 58  &  35.58\%  \\
	    \midrule
	    Nested Loops        & 25  & 29.41\%  & 18  &  11.04\%  \\
        Pointer Operation  & 17  & 20.00\%  & 4   &  2.45\% \\
        \midrule
        Others/Basic       & 14  & 16.47\%  & 80  &  49.07\%  \\
        \midrule
        Total Features     & 85  & -        & 163 & - \\
        \bottomrule
    
    \end{tabular}
	}
\vspace{-5mm}
\end{table}

The statistical results are shown in Table~\ref{tbl-complex-features}. Note that a program can have multiple defined categories, for example, a program can have both multiple loops and nested loops associated with it. We can find that the values for Nested Loops and Pointer Operation in the incorrect samples are higher than these values in the correct samples, which proves that these program features are not learnt well. Furthermore, the number of samples with the simple structure i.e., Others/Basic category in the incorrect samples are far less than the samples in the correct samples, which illustrates that complex program structure is still a challenge for the neural network to learn semantics. Lastly, we cannot claim the learning capacity of the neural network on Do-While and Multiple Loops, since the values of the correct and incorrect samples are near.
\subsection{Threats to Validity}
\label{sec-threats}
\noindent \textbf{Evaluation Tasks.} There are many other code-related tasks such as code translation~\cite{lachaux2020unsupervised}, code summarization~\cite{liu2020retrieval}, code search~\cite{gu2018deep, liu2021graphsearchnet}. We only consider the classification tasks on C programming language since classification tasks are more suitable for quantitative analysis as compared to the generation tasks such as code summarization. Our study provide ideas on how to evaluate program semantics on complicated tasks. 

\noindent \textbf{State-of-the-art Results.} We did not compare with the state-of-the-art results for each task, since these high-performing approaches tend to design more complicated architecture than these basic networks. For example, to achieve the best performance, Devign~\cite{devign} utilized a convolution module to further improve the learnt representations by GGNN. To reduce the complexity of the analysis, we target basic models and conduct a systematic study to explore different categories of code representations, which we believe is fundamental and meaningful.

\noindent \textbf{Hyper-parameter Tuning.} Hyper-parameters affect the performance of each model. For fairness in our study, we tune the hyper-parameters such as embedding dimension, batch size to obtain the best result for each evaluating task. We tune code classification based on the best accuracy, vulnerability detection, and clone detection based on the best f1. {We set the same seed for all experiments for the reproduction and eliminating the potential bias of randomness.}

\section{Related Works}
\label{sec-related}

\noindent \textbf{Source Code Representations.} Feature-based representations~\cite{tfidf1, tfidf2} are commonly used in software engineering tasks as word frequency are important in identifying source code. Other feature-based representations, such as FLUCCS~\cite{fluccs} require domain experts into identifying key features of source code. Sequence-based representation employs distributed representation in learning each token. Many works employ sequence-based representation in clone detection~\cite{cc1}, vulnerability detection~\cite{vd1}, auto-patch identification~\cite{zhou2021spi}, code review~\cite{siow2020core}. These works have proven that sequence-based representations are capable of representing semantic and syntactic information of programs. Researchers investigate deeper representation by traversing the tree-based representation of the source code, such as AST~\cite{code2vec, code2seq, liu2019atom, astnn}. Code2vec~\cite{code2vec} and Code2seq~\cite{code2seq} find an efficient representation of source code by traversing AST and learning on the relevant paths. Many works~\cite{ijcai2017-0423} also employ AST in learning code clones as clones inherently have similar program constructs. While these representations contains the syntactic information of the source code, semantic information is lacking in them as control flow and data flow are not well-model into them.  More works start to look into graph-based representation for programs and their application~\cite{liu2020unified, liu2020retrieval, liu2021graphsearchnet, learning_programs, fernandes2018structured} on GNNs. Miltiadis et al.~\cite{learning_programs} combines the AST node with control flows and data dependency to learn source code graph representations. Wang et al.~\cite{gmn} added data-flow information to AST to improve the performance of clone detection. These works focus on embedding source code into a vector space without losing semantic and syntactic knowledge. Kang et al.~\cite{code2vecsurvey} conduct an code2vec assessment on downstream tasks. It is similar to our work in nature, however, we employ a complicated graph representation of source code and extensively evaluate them.  


\noindent \textbf{Relevant Works to Selected Tasks.} Classifying source code by functionality enables developers to locate functions. Many works~\cite{astnn, tbcnn} focus on classifying source code. Clone detection is a popular topic in software engineering. Several works~\cite{isceclone1, astnn, 10.1109/ASE.2019.00099, code2vecsurvey} relate to clone detection. Vulnerable functions often are not straightforward and contain hidden semantics across multiple functions. Different ways are employed in finding vulnerabilities, such as fuzzing~\cite{10.1145/3236024.3264593, 10.1145/3106237.3106295, 10.1145/3338906.3338975, 274701, issre2021vallnut}, malware detection~\cite{tifs2018spread}, program metrics~\cite{leopard} and deep learning~\cite{devign}. 



\section{Conclusion}
\label{sec-con}
We conduct a systematic study to investigate program representations i.e., Feature-based, Sequence-based, Tree-based, and Graph-based representation across three diverse and popular code-related tasks i.e., code classification, clone detection, and vulnerability detection on public benchmarks. We conclude our findings as follows: (1) Graph-based representation outperforms other techniques by a significant margin. (2) The node type and node textual information are both beneficial in the tree-based and graph-based representation to learn the program semantics, however, node textual information is more critical. (3) Different task relies on task-specific semantics to achieve the best performance, however, a composite graph with the comprehensive program semantics can still yield promising results. By our study, we hope to provide several insights and follow-up directions in the program representation field.
\section{Acknowledgments}
This research is partially supported by the National Key R\&D Programmes of China (No. 2019AAA0104301), National Natural Science Foundation of China (No. 61902395), National Research Foundation, Singapore under its the AI Singapore Programme (AISG2-RP-2020-019), the National Research Foundation, Prime Ministers Office, Singapore under its National Cybersecurity R\&D Program (Award No. NRF2018NCR-NCR005-0001), NRF Investigatorship NRF-NRFI06-2020-0001,  the National Research Foundation through its National Satellite of Excellence in Trustworthy Software Systems (NSOE-TSS) project under the National Cybersecurity R\&D (NCR) Grant award no. NRF2018NCR-NSOE003-0001.

\newpage
\bibliographystyle{IEEEtran}
\bibliography{ref}

\end{document}